\documentclass[journal=nalefd]{achemso}

\usepackage{graphicx}
\usepackage{amsmath,amsfonts,amssymb} 
\usepackage[colorlinks=true,linkcolor=blue,citecolor=blue]{hyperref}
\usepackage{CJK}
\usepackage{siunitx}
\usepackage{xcolor}
\usepackage{endnotes}

\usepackage{caption}

\setkeys{acs}{articletitle = true}

\title{Transport Through a Network of Topological Channels in Twisted Bilayer Graphene} 

\author{Peter Rickhaus}
\email{peterri@phys.ethz.ch}
\affiliation{Department of Physics, ETH Z\"urich, Otto-Stern-Weg 1, 8093 Z\"urich, Switzerland}
\author{John Wallbank}
\affiliation{Centre for Ecology and Hydrology, Maclean Building, Benson Lane, Crowmarsh Gifford, Wallingford, Oxfordshire, OX10 8BB, UK}
\author{Sergey Slizovskiy}
\affiliation{NRC ``Kurchatov Institute'' PNPI,  Gatchina, 188300, Russia}
\author{Riccardo Pisoni}
\affiliation{Department of Physics, ETH Z\"urich, Otto-Stern-Weg 1, 8093 Z\"urich, Switzerland}
\author{Hiske Overweg}
\affiliation{Department of Physics, ETH Z\"urich, Otto-Stern-Weg 1, 8093 Z\"urich, Switzerland}
\author{Yongjin Lee}
\affiliation{Department of Physics, ETH Z\"urich, Otto-Stern-Weg 1, 8093 Z\"urich, Switzerland}
\author{Marius Eich}
\affiliation{Department of Physics, ETH Z\"urich, Otto-Stern-Weg 1, 8093 Z\"urich, Switzerland}
\author{Ming-Hao Liu}
\affiliation{Department of Physics, National Cheng Kung University, Tainan 70101, Taiwan}
\author{Kenji Watanabe}
\affiliation{National Institute for Material Science, 1-1 Namiki, Tsukuba 305-0044, Japan}
\author{Takashi Taniguchi}
\affiliation{National Institute for Material Science, 1-1 Namiki, Tsukuba 305-0044, Japan}
\author{Thomas Ihn}
\affiliation{Department of Physics, ETH Z\"urich, Otto-Stern-Weg 1, 8093 Z\"urich, Switzerland}
\author{Klaus Ensslin}
\affiliation{Department of Physics, ETH Z\"urich, Otto-Stern-Weg 1, 8093 Z\"urich, Switzerland}

\date{\today}

\begin{document}
\begin{CJK*}{Bg5}{bsmi}



\begin{abstract}
 We explore a network of electronic quantum valley Hall (QVH) states in the moir\'e crystal of minimally twisted bilayer graphene. In our transport measurements we observe Fabry-P\'erot and Aharanov-Bohm oscillations which are robust in magnetic fields ranging from $0$ to $\SI{8}{T}$,  in strong contrast to more conventional 2D systems  where trajectories in the bulk are bent by the Lorentz force. This persistence in magnetic field and the linear spacing in density indicate that charge carriers in the bulk flow in topologically protected, one dimensional channels. With this work we demonstrate coherent electronic transport in a lattice of topologically protected states.
\end{abstract}

\textbf{Keywords:}  Twisted bilayer graphene; Topological network; Fabry-P\'erot, Valleytronics; Moir\'e superlattice; Quantum valley Hall effect

\maketitle
\end{CJK*}


\begin{figure}[ht!]
	\centering
	\includegraphics[width=1\textwidth]{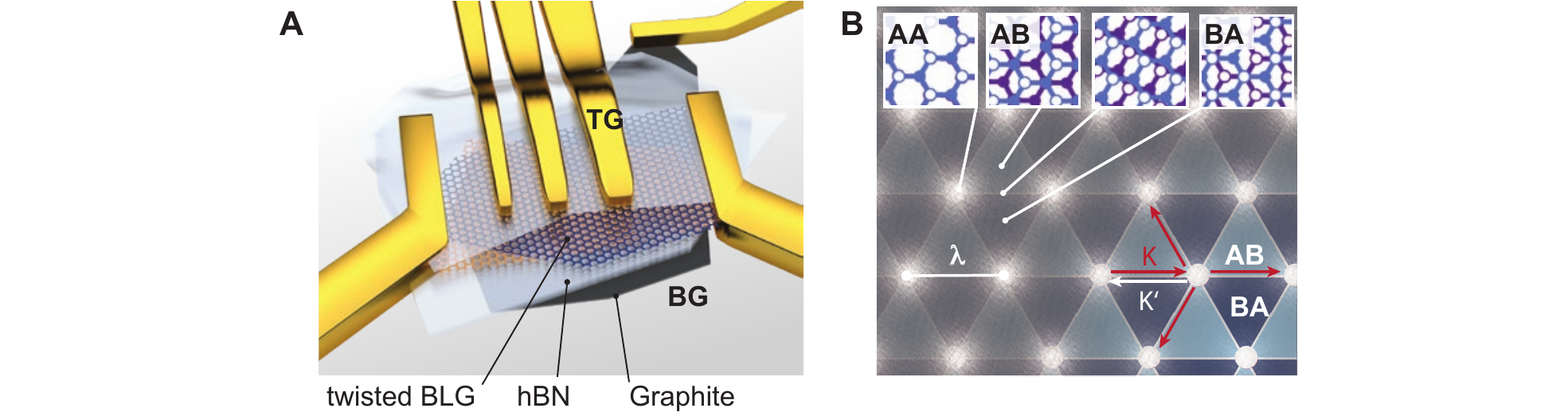}
	\caption{
		\textbf{Schematics of the measured device. A}, twisted bilayer graphene (tBLG) is encapsulated in boron-nitride (hBN). Using the BG, the global density $n_{\rm{out}}$ can be changed, while the TGs allow to change $n_{\rm{in}}$ and $D$. 
		\textbf{B}, Two hexagonal lattices are twisted by $\theta=\SI{0.42}{\degree}$, giving rise to a moir\'e periodicity of $\lambda=\SI{33}{nm}$. By depleting AB and BA regions, helical currents connecting the AA points, arise.
		\label{fig:tBLG1}}
\end{figure}

Topological channels\cite{Kane2005,Konig2007,Young2014,Sau2010}  hold  promises for  quantum computation with reduced decoherence.
 In order to create topological states in  bilayer graphene (BLG), a large displacement field $D$ has to be applied between the two layers. {By this a band-gap opens around the charge neutrality point.} The geometric boundaries at which helical states then emerge are given by stacking faults \cite{Yao2009, Zhang2011, Zhang2013, Hattendorf2013, Vaezi2013, Alden2013, Ju2015, Yin2016}, a smooth transition between AB and BA stacking regions \cite{Wright2011} or the local inversion of $D$ \cite{Zarenia2011,Li2016,Lee2017}. 
 In a moir\'e crystal of twisted bilayer graphene (tBLG), alternating regions of AB and BA stacking naturally exist and they form a superlattice \cite{San-Jose2013}. The AB and BA regions can be depleted by applying large $D$ and the emerging states form a network\cite{San-Jose2013,Andelkovic2018}, as recently shown by STM measurements\cite{Huang2018}. {This network forms due to different valley Chern numbers in the AB and BA stacking regimes.
The condition for its formation is} that the twist angle is sufficiently small, such that the size of the AB/BA regions is large.
{First theories suggested that twist angles $\theta < \SI{0.3}{\degree}$ are required\cite{San-Jose2013}, however, elastic deformations stabilize and enlarge the AB/BA regions and therefore relax this condition\cite{Andelkovic2018}.}
This is in contrast to the emergence of superconductivity \cite{Cao2018a}, which requires a magical twist angle {around $\SI{1}{\degree}$}. 
{Compared to other helical systems, the topological currents flow predominately in the bulk of the sample. This brings the advantage that the system is less sensitive to impurities that originate from processing the sample edge (e.g. in InAs/GaSb systems \cite{Mueller2017})}.

We probe {the topological network using a  Fabry-P\'erot cavity}, formed by a backgate (BG) and a local topgate (TG), and measure charge carrier transmission in a linear conductance experiment. Interfaces between bulk and cavity are semi-transparent, leading to standing waves\cite{Liang2001,Ji2003,Young2009}. {We observe magneto-conductance oscillations that are tuned by density $n$ (Fabry-P\'erot resonances) and magnetic field $B$ (Aharanov-Bohm resonances). The Fabry-P\'erot resonances, at $n$ close to zero},  are periodic in $n$ (rather than $\sqrt{n}$) demonstrating the 1D (rather than 2D) nature of the corresponding channels. Upon application of $B$, Aharanov-Bohm oscillations arise with characteristic areas much smaller than the cavity size but also much larger than the moir\'e unit cell. We find that the characteristic orbits are in the cavity bulk, encompassing several unit cells. In other systems, Fabry-P\'erot resonances are typically suppressed once the cyclotron diameter becomes comparable to relevant device dimensions\cite{Rickhaus2015,Lee2016a}. In our experiments they persist up to $B=\SI{8}{T}$ where the magnetic length ($\SI{9}{nm}$) is much smaller than any device dimension. {The fact that oscillations nonetheless persist indicates that time reversal symmetry cannot be broken or that there is another protective symmetry at play. This hints at topological protection of corresponding 1D states.} Our claims are substantiated by band structure calculations.

The measured device is schematically drawn in Fig.~\ref{fig:tBLG1}A (details in Fig.~S1). tBLG is encapsulated in hBN and contacted with Cr/Au\cite{Wang2013}. The bulk carrier density $n_{\rm{out}}$ can be adjusted using a BG \cite{Overweg2017a}. Three TGs having lithographic lengths $L=200,\,300,\,\SI{400}{nm}$, allow to adjust cavity density $n_{\rm{in}}$ and  displacement field $D$. {$n_{\rm{in}}$, is tuned by the voltages on the topgate and on the graphite backgate, $V_{\rm{tg}}$ and $V_{\rm{bg}}$ respectively, according to the equation: $n_{\rm{in}} = (C_{\rm{tg}}V_{\rm{tg}}+C_{\rm{bg}}V_{\rm{bg}})/e$. The capacitances per unit area are determined from a parallel plate capacitor model, i.e. $C_{\rm{tg}}=\epsilon_0\epsilon_{\rm{hBN}}/d_{\rm{top}}$ and $C_{\rm{bg}}=\epsilon_0\epsilon_{\rm{hBN}}/d_{\rm{bottom}}$, where we use $\epsilon_{\rm{hBN}}=3.2$ and the hBN thicknesses $d_{\rm{top}}=\SI{27}{nm}$, $d_{\rm{bottom}}=\SI{45}{nm}$. To determine the displacement field, we use the simple approximation $D=(D_{\rm{top}}-D_{\rm{bottom})}/2$ and $D_{\rm{top}}=\epsilon_rV_{\rm{TG}}/d_{\rm{top}}$.} The tBLG flake is etched to $W=\SI{4.6}{\mu m}$ giving cavities with $L\ll W$. Therefore, many parallel channels follow the same interference condition and standing waves in transport direction dominate the conductance.

The small twist angle is obtained by tearing a large graphene flake in the middle and picking up one half. The remaining part is twisted by $\theta=\SI{0.5}{\degree}$ and also picked up, following the procedure described in references \cite{Kim2016,Cao2016}. 
{It is this careful fabrication that guarantees a well controlled and homogeneous moir\'e periodicity (a detailed description is given in the supplementary information Fig.~S2).}

Conductance measurements are performed using a standard low-frequency lock-in technique at $\SI{1.5}{K}$.
From the Hofstadter butterfly pattern (Fig.~S3) we extract a density $|n_2|\approx\SI{0.8e12}{cm^{-2}}$ at which the first band is completely filled\cite{Kim2017}, corresponding to $\theta=\SI{0.42}{\degree}$.
In Fig.~\ref{fig:tBLG1}B, two hexagonal lattices, twisted by $\SI{0.42}{\degree}$ are shown, exhibiting a large period moir\'e superlattice. 

\begin{figure*}[ht!]
	\centering
	\includegraphics[width=1\textwidth]{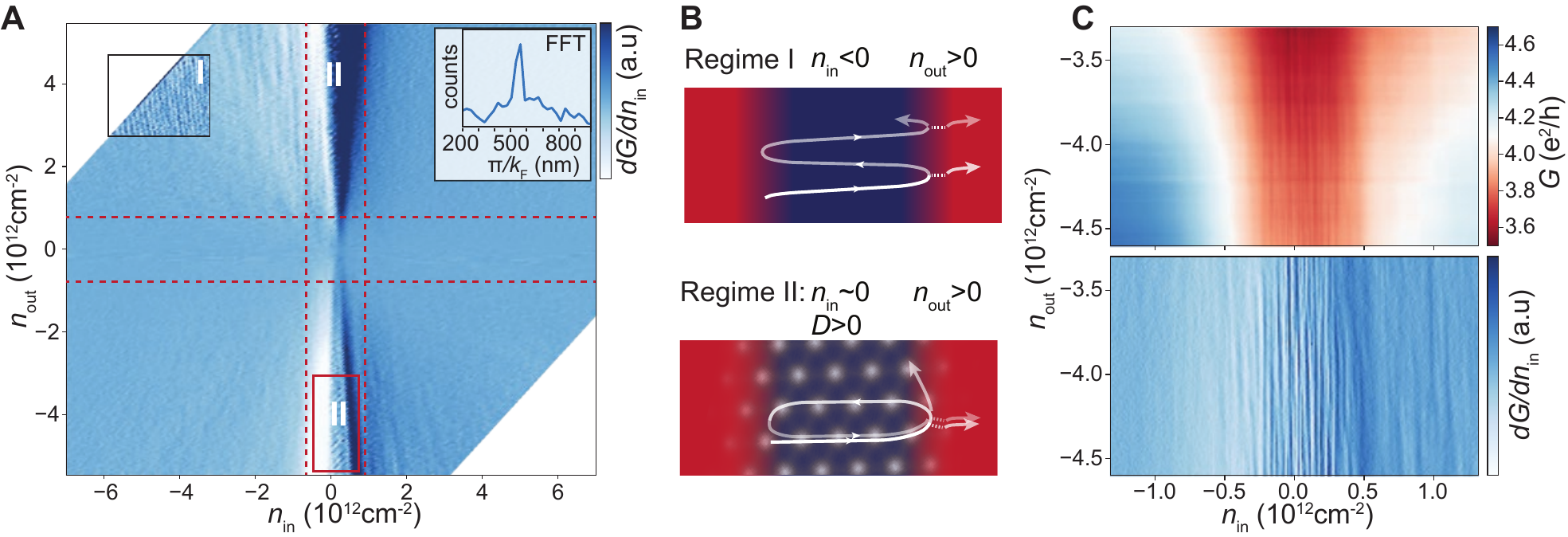}
	\caption{
		\textbf{Fabry-P\'erot (FP) oscillations measured at $\SI{1.5}{K}$. A},
		Differential conductance $dG/dn_{\rm{in}}(n_{\rm{in}},n_{\rm{out}})$. Red dashed lines denote the density $n_2$. In regime I, FP oscillations appear for $n_{\rm{in}}\ll 0$ and $n_{\rm{out}}\gg 0$, as can be seen in the high-resolution scan in the marked window. The corresponding FFT (averaged over $n_{\rm{out}}$) is shown in the inset.
		\textbf{B}, FP resonances are due to standing waves in a cavity formed by the topgate. In regime II, the presence of topological channels is expected. 
		\textbf{C}, A zoom into regime II (marked with a red solid square in \textbf{A}) is shown (above: $G$, below: $dG/dn_{\rm{in}}$).
		\label{fig:tBLG2}
	}
\end{figure*}

In the measurement $dG/dn_{\rm{in}}(n_{\rm{in}},n_{\rm{out}})$ (Fig.~\ref{fig:tBLG2}A), $|n_2|$ is marked with red dashed lines. We first focus on the bipolar n-p-n regime I, {where the densities $n_{\rm{in}}$, $n_{\rm{out}}$ are large but have opposite signs (a negative sign is used for charge carriers occurring at energies smaller zero). The displacement field can, but does not have to be large in this regime}. For a semi-transparent interface, standing waves form as sketched in Fig.~\ref{fig:tBLG2}B, following the 2D-Fabry-P\'erot interference condition $2L=j\cdot2\pi/k_{\rm{F}}$ where $j=1,2,...$ and $k_{\rm{F}}\approx\sqrt{n\pi}$. The observed pattern is very similar to measurements in mono-\cite{Young2009,Rickhaus2013,Handschin2017a} and bilayer graphene\cite{Varlet2014}. The extracted cavity length $L=\SI{550}{nm}$ (see inset), is larger than the designed $L=\SI{400}{nm}$. This discrepancy is due to the smooth transition between cavity and bulk and is analyzed in detail in the supplementary material of reference \cite{Handschin2017a}. {The observation of standard Fabry-P\'erot oscillations in regime I shows that ballistic cavities with standing waves form.}

We now focus on oscillations at small $n_{\rm{in}}<n_2$ and large D (regime II, Fig.~\ref{fig:tBLG2}C). These resonances occur in a regime where we expect that the AB/BA regions are depleted and the super-lattice symmetries affect the behavior in $B$.

\begin{figure*}[ht!]
	\centering
	\includegraphics[width=1\textwidth]{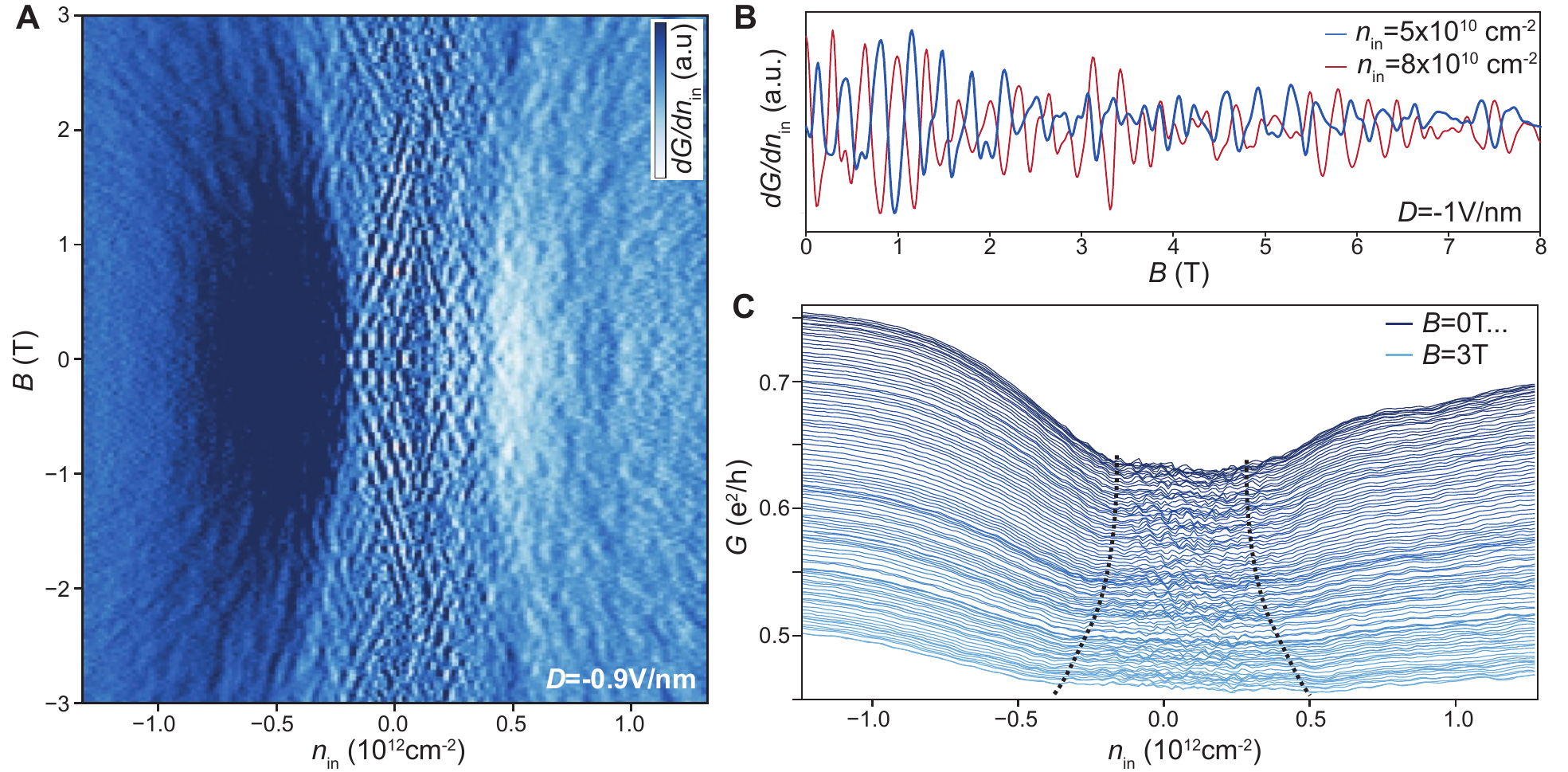}
	\caption{
		\textbf{Magneto-conductance oscillations. A}, A crossed resonance pattern emerges from oscillations at low $n_{\rm{in}}$ and large $D$.
		\textbf{B}, Two lines $dG/dn_{\rm{in}}(B)$ for fixed $n_{\rm{in}}$ with an extended B-field range. Maxima and minima alternate up to $B=\SI{8}{T}$.
		\textbf{C}, $G(n_{\rm{in}})$ traces reveal that the background-conductance is nearly independent of $n_{\rm{in}}$.
		\label{fig:tBLG3}}
\end{figure*}

In a perpendicular magnetic field, trajectories of charge carriers, bouncing between two semi-transparent mirrors, bend due to the Lorentz force. Standard Fabry-P\'erot oscillations require a cyclotron diameter $2R_c$ \cite{Rickhaus2015}:
\begin{equation}\label{eq:Rc}
2R_{\rm{c}}=2\frac{\hbar k_{\rm{F}}}{eB}>L
\end{equation}
In our system, this condition holds true in regime I (see Fig.~S4), where oscillations have vanished for $B>\SI{0.5}{T}$, but not in regime II. There, the corresponding magnetoconductance map (Fig.~\ref{fig:tBLG3}A) reveals a periodic pattern of crossed resonances, evolving continuously from $0$ to $\pm\SI{3}{T}$. {The crossed pattern is formed by diagonal lines of opposite slope in the $n-B$-plane.} For the given density range, $2R_{\rm{c}}\geq L$ for $B\lesssim\SI{0.4}{T}$, the resonance pattern apparently neither disappears nor changes at $\SI{0.4}{T}$, but persists up to at least $\SI{8}{T}$ as seen in Fig.~\ref{fig:tBLG3}B where we depict two traces $dG/dn_{\rm{in}}(B)$ for slightly different values of $n_{\rm{in}}$ (for clarity, a smoothened background has been removed). Up to a magnetic field of $\SI{8}{T}$ (and presumably beyond, $\SI{8}{T}$ was the maximum available field in our cryostat) the maxima and minima alternate periodically.

{In other two-dimensional systems, resonances that depend on $B$ and $n$ at high magnetic fields (i.e. in the quantum Hall regime) were attributed to either single-electron charging of Landau levels in confined geometries or to Aharanov-Bohm interferences.\cite{Zhang2009a,Overweg2018} These two effects are distinguished by the sign of slope in the $n-B$-plane.\cite{Zhang2009a}. In our measurements, oscillations display both, positive and negative slopes simultaneously (Fig.~\ref{fig:tBLG3}A) and are therefore inconsistent with electron charging as a possible origin. For the case of Aharanov-Bohm oscillations however, a crossed pattern can be explained if the corresponding area is encircled both clock- and counterclockwise.  In contrast to the above mentioned measurements\cite{Zhang2009a,Overweg2018}, the resonances persist from the Quantum Hall regime down to low magnetic fields (Fig.~\ref{fig:tBLG3}A,B), and are thus not linked to the existence of Quantum Hall edge channels. This is a strong indication that the charge carriers already flow in one-dimensional channels for all magnetic field considered such that their trajectories remain unaffected by the magnetic field. Another, yet weaker, indication for one-dimensional transport is seen from the conductance traces (Fig.~\ref{fig:tBLG3}C) which are rather flat in the regime where the AB/BA regions are gapped (marked with dashed borders). This indicates that the number of conducting channels does not change with $n_{\rm{in}}$ which is again consistent with a fixed number of one-dimensional channels.}

{More quantitative information can be obtained from the resonance periods in magnetic field and density. These  are linked to the encircled area $A$ and the total length $L_{\rm{tot}}$  of the coherent trajectories by the Bohr-Sommerfeld resonance condition: }	
\begin{equation}\label{eq:Interferencecondtion} 	
	j= L_{\rm{tot}} \frac{k_{\rm{F}}}{2\pi} \pm A\frac{B}{\phi_0}
\end{equation} where $j$ is an integer and $\phi_0=h/e$. {The spacing between two maxima is then given by $j-(j-1)=L_{\rm{tot}} \frac{\Delta k_{\rm{F}}}{2\pi} \pm A\frac{\Delta B}{\phi_0}$. From $\Delta B=\SI{0.37}{T}$ (extracted from Fig.~\ref{fig:tBLG3}B) we obtain  $A=\phi_0/\Delta B=\SI{11200}{nm^2}$.} 
{This area is much larger than the area of a moir\'e unit cell, i.e. $\approx\SI{950}{nm^2}$}. On the other hand, the entire area of the top-gated cavity is $L\cdot W\approx\SI{2e6}{nm^2}$ which is two orders of magnitude too large. Consequently, the interfering paths must be located in the cavity bulk.

{By analyzing the spacing in density, $\Delta n_{\rm{in}}$, we can extract information about the length $L_{\rm{tot}}=2\pi/\Delta k_{\rm{F}}$ of the interference path. Importantly, $k_{\rm{F}}\sim n_{\rm{in}}$ (not $k_{\rm{F}}\sim\sqrt{n_{\rm{in}}}$)  since charge carriers flow in one-dimensional (not two-dimensional) channels. This leads to resonances following \textit{diagonal} lines in the $n_{\rm{in}}$-$B$-plane ($k_{\rm{F}}\sim\sqrt{n_{\rm{in}}}$ would lead to parabolic lines in the magnetoconductance map, which is not observed). To convert $n_{\rm{in}}$, which is the (two-dimensional) density in the twisted bilayer graphene flake tuned by the gate voltages, into a one-dimensional density $n_{\rm{1D}}$ we divide by the number of channels  per unit area, $N_{\rm{ch}}=2\sqrt{3}/\lambda$ (for details see supporting information, Eq.~4). To do so we use the moir\'e periodicity $\lambda=\SI{33}{nm}$ obtained from the Hofstadter butterfly, Fig.~S3. For $\Delta n_{\rm{in}}=\SI{4.7e10}{cm^{-2}}$ and using $L_{\rm{tot}}=2\pi/\Delta k_{\rm{F}}=8\sqrt{3}/(\lambda\Delta n_{\rm{in}})$ we obtain $L_{\rm{tot}}\approx\SI{870}{nm}$.}

{The extracted area and circumference correspond  to  trajectories that encircle a long and narrow object. For a rectangle, it is straightforward to calculate the corresponding length $\tilde{L}=\SI{408}{\nm}$ and width $\tilde{w}=\SI{27}{nm}$. We note here that these values are close to the designed cavity length $L=\SI{400}{nm}$ and the height of the moir\'e unit cell $\lambda\sqrt{3}/2=\SI{29}{nm}$ which also corresponds to the shortest distance between two topological channels. This suggests that one row of AB/BA regions is encircled. However, also other trajectories  are possible. In the topological network there are three valley-preserving scattering possibilities (red arrows in Fig.~\ref{fig:tBLG1}B) at every 'node' (AA stacking region). This allows for large and complex paths in the network. Especially, paths that do not require intervalley scattering (see discussion in Fig.~S7B) are possible. However, closed trajectories consistent with the extracted area and length are long and narrow and if they do connect the two cavity interfaces then the cavity length $L$ is an important parameter. Since $A/L\approx\lambda\sqrt{3}/2$ and $L_{\rm{tot}}-2L\approx2\lambda$ this is the only kind of trajectory that is consistent with our experimental results.}
	
\begin{figure*}[ht!]
	\centering
	\includegraphics[width=1\textwidth]{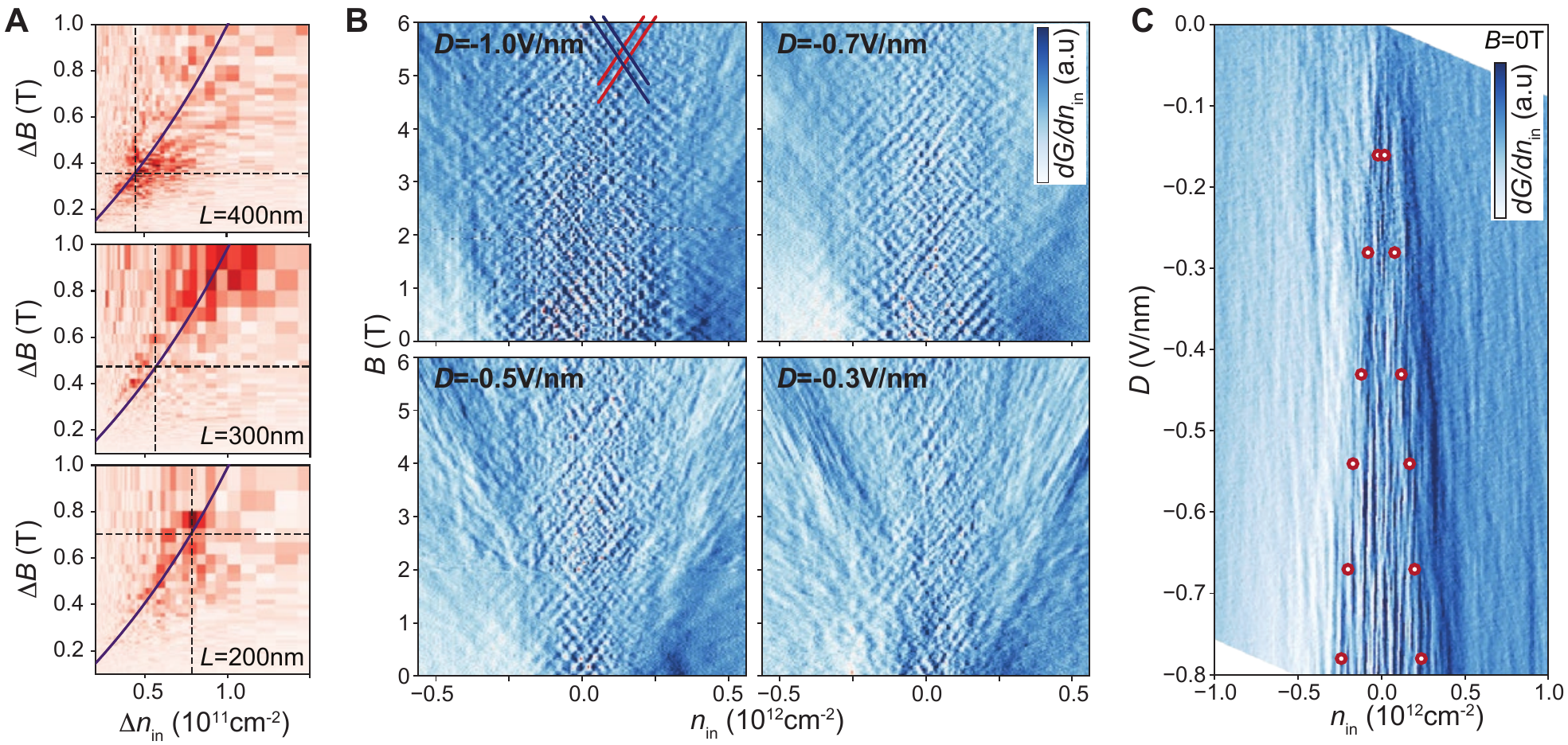}
	\caption{
		\textbf{Dependence on $L$ and $D$. A}, 2D FFT of magneto-conductance maps for $D=\SI{-1}{V/nm}$ and  $L=200,300,\SI{400}{nm}$. The solid line shows the Bohr-Sommerfeld quantization for trajectories with varying length encircling one row of AB/BA regions. The dashed lines depict the expected $\Delta B$ and $\Delta n$  for the designed $L$.
		\textbf{B}, $dG/dn_{\rm{in}}$ for decreasing $D$ and $L=\SI{400}{nm}$. The $n_{\rm{in}}$ range where the interferences are observed is shrinking. 
		\textbf{C}, Interference pattern as a function of $D$ and $n_{\rm{in}}$. Red dots mark the  theoretically expected boundary for the resonances.
		\label{fig:tBLG4}}
\end{figure*}

{By measuring the crossed resonance pattern with different topgates, it is possible to see how (and if) the resonance pattern depends on the designed cavity length $L$. In Fig.~\ref{fig:tBLG4}A we show the results of a 2D fast Fourier transform (FFT) of magneto-conductance oscillations for cavities with $L=400,300,\SI{200}{nm}$. Apparently, the oscillations have a strong dependence on $L$ and become slower in both $\Delta B$ and $\Delta n_{\rm{in}}$ for decreasing $L$, meaning that both the encircled area $A$ and  circumference decrease in size. As a guide to the eye we depict the value of $\Delta B$ and $\Delta n_{\rm{in}}$, that we would expect for a given $L$ by assuming that one row of AB/BA regions is encircled, with dashed lines. The solid line is for arbitrary $L$ (details are given in the supplementary material). The measurements for $\Delta B$ and $\Delta n_{\rm{in}}$ for different cavities appear to be consistent with straight parallel trajectories encircling one row of AB/BA regions. Even though the states below the $\SI{300}{nm}$-sized topgate seem to resonate in an effectively shorter cavity, their trajectories also seem to encircle one row of AB/BA regions as can be seen from the agreement with the solid line.  
}

Finally we discuss the dependence on $D$. The measurements in Fig.~\ref{fig:tBLG4}B,C show that the resonance-pattern boundaries move closer when decreasing $D$. Such behavior is expected within the topological model. By lowering $D$, the induced gap size $\Delta$ (values in Fig.~S9) in the AB/BA regions shrinks and these regions start to become populated already at lower $n_{\rm{in}}$. 
{At the boundary of the resonance pattern, the Fermi-surface looses its one-dimensional character, as indicated by smearing of Fermi velocities projected onto the direction of 1D channel (supporting information Fig.~S10B). This leads to dephasing and smearing of the interference pattern (see Fig.~S10C). In Fig.~\ref{fig:tBLG4}C the densities, where the calculated Fermi velocity smears strongly, are marked with red dots, providing good agreement with the experimental data.}


\paragraph*{Conclusion}
We have fabricated tBLG with {a twist angle of} $\theta\approx\SI{0.4}{\degree}$. We measured {a crossed interference pattern which we explained by Aharanov-Bohm and Fabry-P\'erot oscillations of trajectories that encircle an area clock- and anti-clockwise. The interference pattern persists from zero to large magnetic fields, which indicates that the charge carriers flow in one-dimensional channels. From the oscillation period in density and field we calculated area and circumference of the (dominant) resonant paths and found that their length is comparable to the cavity length and the width to the moir\'e periodicity. Similar loops are found for different gate lengths. The range (in density) within which the oscillations can be observed exhibits a dependence on displacement field that is consistent with the opening of a gap in AB and BA regions of the twisted bilayer graphene. Our observations are good indications that electrons form coherent paths within a network of topological channels that originates from the moir\'e superlattice.}

Networks of helical channels offer several advantages for topologically protected quantum states: The two-dimensional nature of the network allows to perform complex valleytronic operations and, as demonstrated in this work, stabilizes coherent bulk transport phenomena such as Fabry-P\'erot oscillations in magnetic field and against disorder. Furthermore, avoiding the physical edge leads to a better defined environment which improves topological protection. Our carbon based system is flexible, making it an important building block for scalable and protected valleytronic devices.

\section*{Acknowledgments}

We thank Lucian Covaci, Beat Braem, Andreas Baumgartner and Lujun Wang for fruitful discussions and Peter M\"arki, Erwin Studer and FIRST staff for their technical support.\\
 \textbf{Funding:} We acknowledge financial support from the European Graphene Flagship, the Swiss National Science Foundation via NCCR Quantum Science and Technology (QSIT) and ETH Z\"urich via the ETH fellowship program and the Taiwan Minister of Science and Technology (MOST) under Grant No. 107-2112-M-006 -004 -MY3. Growth of hexagonal boron nitride crystals was supported by the Elemental Strategy Initiative conducted by the MEXT, Japan and JSPS KAKENHI Grant Number JP15K21722. \\
 \textbf{Authors contributions:} PR fabricated the devices and performed the measurements. JW calculated the band structure. SS and M-HL helped to develop the theoretical understanding. SS and PR derived eq. 5,8 in SI. SS calculated gap size, Fermi-velocities and the parameter b (eq. 9,10). RP, HO, YL and ME supported device fabrication and data analysis. KW and TT provided high-quality Boron-Nitride. KE and TI supervised the work.\\
  \textbf{Competing interests:} Authors have no competing interests\\
  \textbf{Data and materials availability:} All data is available in numerical form upon request.\\
  \\
\textbf{Supporting Information:}\\
Details on device design\\
 Relative orientation of the moir\'e lattice and the TG/BG interface\\
Device fabrication and homogeneity of the moir\'e pattern\\
Hofstadter butterfly\\
Fabry-P\'erot resonances of region II in $B$\\
Measurement of another sample\\
Calculation of Fabry-P\'erot area and length\\
Model for one row of AB/BA regions\\
Formation of semi-transparent mirrors at the device boundaries\\
Discussion of the band structure\\
Displacement field, bandgap and energy scales\\
Relation of $\Delta B$ with $\Delta n$\\

\bibliographystyle{acs}
\bibliography{2018-tBLG-paper-final}

\end{document}